# NEW FORMS OF STRUCTURE IN ECOSYSTEMS REVEALED WITH THE KURAMOTO MODEL

John Vandermeer[1,2]
Zachary Hajian-Forooshani[1]
Nicholas Medina[1]
Ivette Perfecto[2,3]

1. Department of Ecology and Evolutionary Biology, University of Michigan, Ann Arbor; 2. Program in the Environment, University of Michigan, Ann Arbor; 3, School for Environment and Sustainability, University of Michigan, Ann Arbor;

Corresponding Author:  John Vandermeer, jvander@umich.edu

## ABSTRACT

Ecological systems, as is often noted, are complex.  Equally notable is the generalization that complex systems tend to be oscillatory, whether Huygens' simple patterns of pendulum entrainment or the twisted chaotic orbits of Lorenz' convection rolls. The analytics of oscillators may thus provide insight into the structure of ecological systems. One of the most popular analytical tools for such study is the Kuramoto model of coupled oscillators. Using a well-studied system of pests and their enemies in an agroecosystem, we apply this model as a stylized vision of the dynamics of that real system, to ask whether its actual natural history is reflected in the dynamics of the qualitatively instantiated Kuramoto model.  Emerging from the model is a series of synchrony groups generally corresponding to subnetworks of the natural system, with an overlying chimeric structure, depending on the strength of the inter-oscillator coupling. We conclude that the Kuramoto model presents a novel window through which interesting questions about the structure of ecological systems may emerge.

Definitions of objects to be studied and underlying assumptions about how those objects relate to one another are basic to any science. In particular, the history of ecology can be traced through a succession and accumulation of defined objects, from spatial vegetation patterns in succession (Tansley 1935), to coupled populations as dynamical systems (Lotka 1925; Volterra 1931; Nicholson & Bailey 1935), to energy and matter flows (Lindeman 1942), and recently to individual, population, or species networks of food webs (Borrett et al. 2013). Less well-known is the abstract generalization of Platt and Denmann (1975) that a "… most important characteristic of complex systems [is that] the functional relations between the system components be of the nonlinear kind", and that "A crucial characteristic of nonlinear systems is their disposition toward periodic behavior, even for non-periodic boundary conditions." Thus, in addition to the well-known oscillatory parameter spaces for resource/consumer dynamics, the Platt and Denmann framework suggests that such may be the case for all ecological systems.

If it is true that ecological systems are characteristically periodic, interacting species can be studied as collections of oscillators. Whether predators and their prey, or herbivores and their plants, or parasites and their hosts, consumers eat their resources and then find themselves in a resource-poor environment, thus lowering their growth rate and allowing the resources to recuperate – a fundamentally oscillatory process. To the extent that the consumers tend to overlap in their diets, or the resources interact, these oscillators are coupled with one another (Vandermeer 2004). The collection of traditionally defined objects of ecology (e.g., vegetational patterns, population dynamics, energy flows) thus might be augmented by considering the simple idea of coupled oscillators, a common vehicle for developing theory in many branches of science, from electronics to neurobiology.

Coupling oscillator sets has already gained traction in the ecological literature (e.g. Hastings, 1993; 2001; 2004; 2010; Blasius and Stone, 2000; Earn et al., 1998; Goldwyn and Hastings, 2009; Wall et al., 2013; Satake, and Koizumi, 2008; Vandermeer, 1993; 1994; 2004; 2006), mostly from a theoretical perspective, including the potential to generate chaotic behavior (Hastings and Powell, 1991; Vandermeer, 2006). However, a perhaps more basic question stimulated by the pioneering work of Arthur Winfree (1967), is what will be the patterns of synchrony within the collection of coupled oscillators? The early example of pendulum clocks synchronizing over time due to even weak connections between them is legion (Benicá et al., 2009; Vandermeer, 2006), and begins with the question of whether two oscillators will synchronize with the same phase or with opposite phases. With this emphasis on the phase of the oscillators, Winfree approached the question by interrogating the dynamics of the system from the point of view of the angle defining a point in the cycle of two variables (which is to say the phase of the oscillators). Elaborating on this insight, the model of Kuramoto (1975) has become something of a standard approach to analyzing synchrony in collections of coupled oscillators (Strogatz, 2000).

**The Kuramoto model**

Consider angle $\Theta$ made by a point on the unit circle, to represent the position on the limit cycle of the consumer/resource oscillator (Vandermeer, 1994). Presuming that synchronization will occur, the general Kuramoto model holds that,

$$\frac{d\Theta_i}{dt} = \omega_i + \frac{K}{N}\sum_j (\sin \Theta_i - \sin \Theta_j) \qquad 1$$

where $\omega_i$ is the winding number of oscillator i (the rate of advancement on the circle dictated by the inherent oscillations), K is the intensity of coupling, and N is the number of oscillators. This model is commonly used when the phases of the oscillations are taken to be the key dynamical force. In its classical form, the model assumes all oscillators identical and couplings are taken to be global (all to all).

This elementary model yields a simple and universal result. With random initiations of $\Theta$, and low coupling, no synchrony occurs. As coupling intensity (K) increases, a critical value exists at which point all oscillators rapidly synchronize, and remain synchronized with further increases in coupling intensity. This model has been useful in studying large systems of coupled oscillators (Strogatz, 2000). One interesting result is that among groups of strongly synchronous pairs of oscillators, for some arrangements of coupling, there are individual oscillators that fail to synchronize with the rest, creating a so-called chimeric pattern (Jaros et al., 2015), although for finite N, they are now generally thought to be extremely long transients (Panaggio & Abrams, 2015).

Based on the reality of coupling types in consumer/resource systems, when two consumers share at least one key resource they will experience competition from one another. If their sharing is relatively weak, they will converge on a pattern of relative in-phase synchrony with one another (Vandermeer, 1997). If, contrarily, coupling is through the competitive interactions of the resources, the oscillators will converge on a pattern of relative anti-phase synchrony with one another (Vandermeer, 1999). This arrangement has led to some speculations on its meaning for ecological communities in general (Vandermeer 2008) as well as empirical confirmation in the field (Benica et al., 2009). Elsewhere it has been shown that the pattern of coupling based on the classical consumer resource model of MacArthur (MacArthur, 1984) follows precisely the qualitative predictions of coupling patterns from the Kuramoto model (Hajian-Forooshani and Vandermeer, 2020).

To represent an actual empirical community, we relax the assumption of the universal constant coupling in Kuramoto's model (equation 1), obtaining,

$$\frac{d\Theta_i}{dt} = \omega_i + \frac{k}{N}\sum_j \gamma_{i,j}(\sin\Theta_i - \sin\Theta_j) \quad 2$$

where Kuramoto's mean field approach has been disaggregated with the adjacency matrix $\Gamma$ (with elements $\gamma_{i,j}$) stipulating the coupling of each pair of oscillators. For a non-weighted, non-directed graph, $\gamma_{i,j}$ is either 0 or 1.

The degree of synchrony is frequently measured by Kuramoto's order parameter, which is the absolute value of z where,

$$z = \frac{1}{N}\sum_{j=1}^{N} e^{i\Theta_j}, \quad 3$$

and N is the number of oscillators.

The behavior of the Kuramoto model applied to ecological situations under various assumptions about the distribution of the $\gamma_{i,j}$ has been analyzed, albeit infrequently. For example, Banerjee and colleagues (2016) studied the behavior of the model with $\gamma_{i,j}$ distributed as a distance related power law in a spatially explicit framework, and Girón and colleagues examined the consequences of allowing some of the $\gamma_{i,j}$ to be positive and others negative. In an earlier work Hajian-Forooshani and Vandermeer (2020) compared a six-dimensional predator/prey framework modeled in the Lotka-Volterra style to the same framework in the Kuramoto model showing that the synchronization patterns were qualitatively identical. Numerous studies, not necessarily ecological, have focused on a central feature of the model, the chimeric state that inevitably emerges when the $\gamma_{i,j}$ collectively are too small to engage all the oscillators in complete collective synchrony but too large to permit complete independence of oscillator behavior (Belykh et al., 2016). Others have noted the emergence of subnetworks as synchrony groups (Menara et al., 2109), depending on the distribution of the $\gamma_{i,j}$ for several theoretical distributions.

**A real ecological system**

As interesting as the theoretical diversions of the basic model are, thus far there has not been an attempt at applying the Kuramoto framework to a real ecological system. Here, we utilize the well-documented system of pests and their natural enemies in the coffee agroecosystem (Vandermeer et al., 2019; Perfecto and Vandermeer, 2015) as a system in which predators/parasitoids/diseases attack four well-known pests of coffee, and therefore, can be viewed as a single system of oscillators. The four pests are 1) the coffee berry borer (*Hypothenemus hampei*), 2) the coffee leaf miner (*Leucoptera coffeella*), 3) the green coffee scale (*Coccus viridis*) and 4) the coffee leaf rust fungus (*Hamaelia vastatrix*). We ask, to what extent does the Kuramoto model provide insight into the community structure of this well-defined real community? Will the Kuramoto model reflect what we know about the system? Will it provide any insights regarding the structure of this community? We diagram the basic system in Figure 1. It is critical to note that this rendering, although it includes top predators and parasitoids (those that eat the consumers of the prey), is a simplified rendering, emphasizing the direct energy transfer interactions and ignoring the well-known indirect higher-order interactions (Vandermeer et al., 2019). Anticipated further studies will incorporate those more complex additions of reality.

Applying the model to this well-studied system we seek to determine whether modules of synchronization appear and whether these modules reflect the reality of what we know about that real system. It is not claimed that the Kuramoto model will provide extra evidence that is not apparent from the fairly obvious structure of this simple community. Rather, we seek to 1) use the real system to interrogate the model, and 2) ask whether what we might conclude from the model concords with what we know about the real ecological system.

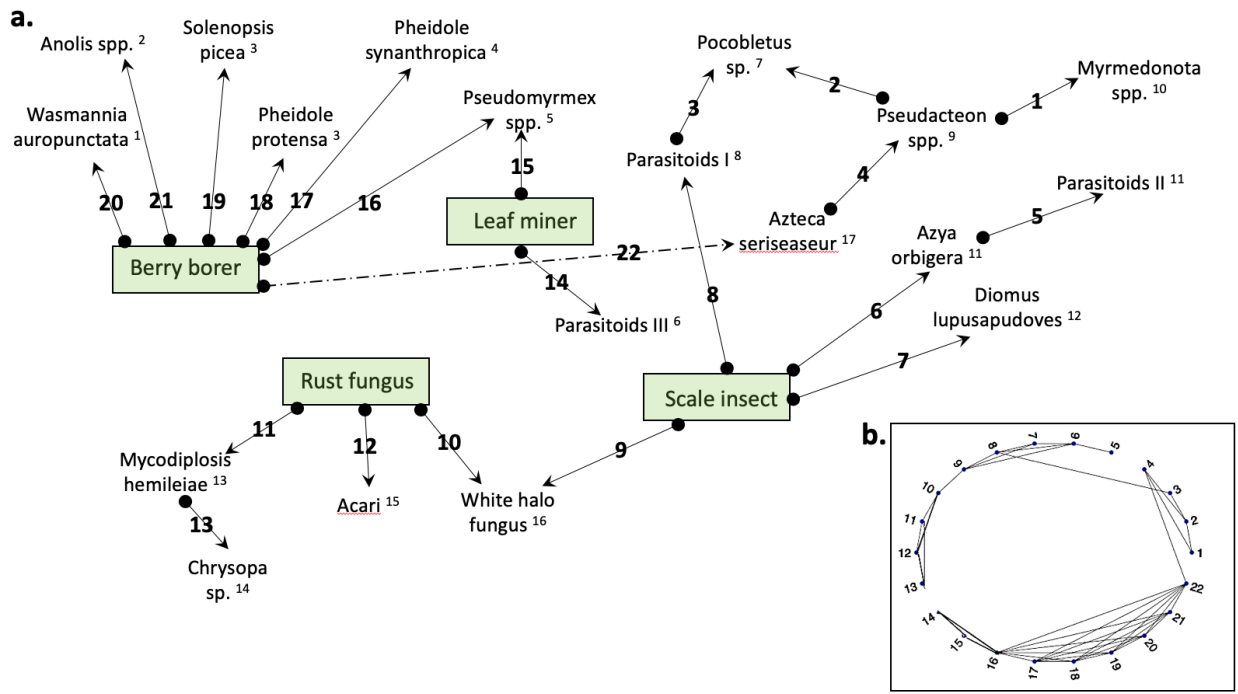

*Figure 1. Diagram of the study system. a. Each arrow represents a negative (closed circle) effect of a consumer (predator or parasitoid or disease) on one or more of the four pests, the berry borer (*Hypothenemus hampei*), the coffee leaf miner (*Leucoptera coffeella*), the scale insect (*Coccus viridis*) or the coffee leaf rust fungus (*Hamaelia vastatrix*), or on one of the consumers of those pests. The connections are thus oscillators and are numbered for reference in bold numbers (a total of 22 oscillators, with one, number 22, illustrated with a dashed line as it provides a key element to the overall network structure of the system (see text). b. Rendering of the network of oscillators and their particular couplings (numbers refer to the number of the oscillator in part a). Footnotes (small superscripts) for each of the organisms refer to:1. De la Mora et al, 2015; Gonthier et al., 2013;Morris and Perfecto, 2016. 2. Monagan et al., 2017. 3.Perfecto and Vandermeer, 2013; Gonthier et al., 2013. 4. Ennis and Philpott, 2017; Jiménez-Soto et al., 2013; 5. Philpott, 2010; Philpott et al., 2008; 2012. 6. De la Mora et al., 2015. 7. Marin et al., 2016. 8. Uno, 2007. 9. Jackson et al., 2012; 2012a; Vandermeer et al., 2009; 2014; 2017. 10.*

**The Kuramoto model and synchrony groups**

The dynamics of coupled ecological oscillators is complicated and strongly dependent on the nature of the coupling (Vandermeer, 1993; 2004; Hajian-Forooshani and Vandermeer, 2000), with in-phase synchrony resulting from predators sharing prey and anti-phase synchrony resulting from prey competing with one another. Empirical verification of these theoretical propositions is rare (e.g., Beninca et al., 2009). However, the network we seek to investigate (Figure. 1) emerges from the "pest" guild associated with the coffee agroecosystem and thus involves only predators consuming prey, and thus are expected, if coupling is weak enough, to produce nearly in-phase synchrony. More complicated networks that include competition at lower trophic levels would require, minimally, the expansion of equation 2 to include both positive and negative K (Girón et al., 2016). Here, all of the oscillators are predator/prey (in principle). Based on the simple idea that oscillators are coupled as long as one element is shared (two predators eating the same prey, two prey eaten by the same predator, or a trophic triplet [chain]) we established an adjacency matrix, which is reflected visually in the graph in the inlay of Figure 1. Using this matrix as $\Gamma$, we employed the extended Kuramoto model at various coupling strengths (with identical winding numbers = 0.01), where every coupling of oscillators (Figure. 1) is at the same strength, K. The general result shows particular patterns of synchrony, occurring in groups of oscillators, what we call 'synchrony groups'. We judge two oscillators, i and j, to be in the same synchrony group if the difference in $\Theta$ for the two oscillators < C, where the critical value of C is a parameter that may be tuned, and indicates membership in a group. For every pair of oscillators we computed,

$$|c_{i,j}| = \sqrt{(sin\Theta_i - sin\Theta_j)^2 + (cos\Theta_i - cos\Theta_j)^2} \qquad 4$$

to compare to C (for all simulations reported herein, C = 0.01 radians). The "stability" of synchronous groups was determined by randomly initiating the model 20 times and recording the number of times (out of 20) that oscillators i and j fell into the same group. If two oscillators fell into the same group at least 19 of the 20 runs, they were judged to be in the same "persistent" synchronous group.

Calculating the order parameter over a range of values of the coupling coefficient, we obtain the usual rapid increase to full coupling (Figure. 2). Contrary to the classic Kuramoto model, we do not see the lower plateau at lower values of the coupling coefficient, nor a critical transition to full coupling, as occurs in the classic model. We attribute this to the complexity of the coupling of the system, as well as its small size (Townsend et al., 2020). With the coupling K=0, a relatively high value of the order parameter is obtained (approximately 0.4 as an average), so any approach to 1 is obscured by the small range (from < 0.1 to 1.0), suggesting that the order parameter is not necessarily an efficient measure of synchronization with such a small number of oscillators.

Contrarily, when we extract only the oscillators associated with an ecologically significant subset of the oscillators, such as the coffee berry borer sub-community (oscillators 17-21, in Figure 1), we obtain a result that mirrors the classic results of the Kuramoto model (Figure 2b). This suggests that key, well-connected modules of interacting oscillators within our network behave qualitatively similarly to the classic Kuramoto model (Figure 2b), but this dynamic is obscured when considering the full network of oscillators (Figure 2a), as is typically done with uniformly global coupling. As with many real networks (Barabasi 1999), most nodes in this community have few links, potentially increasing their ability to synchronize, and increasing the sum that is the real global order parameter. This result highlights the utility of the criterion $|c_{i,j}| < C$, from equation 4.

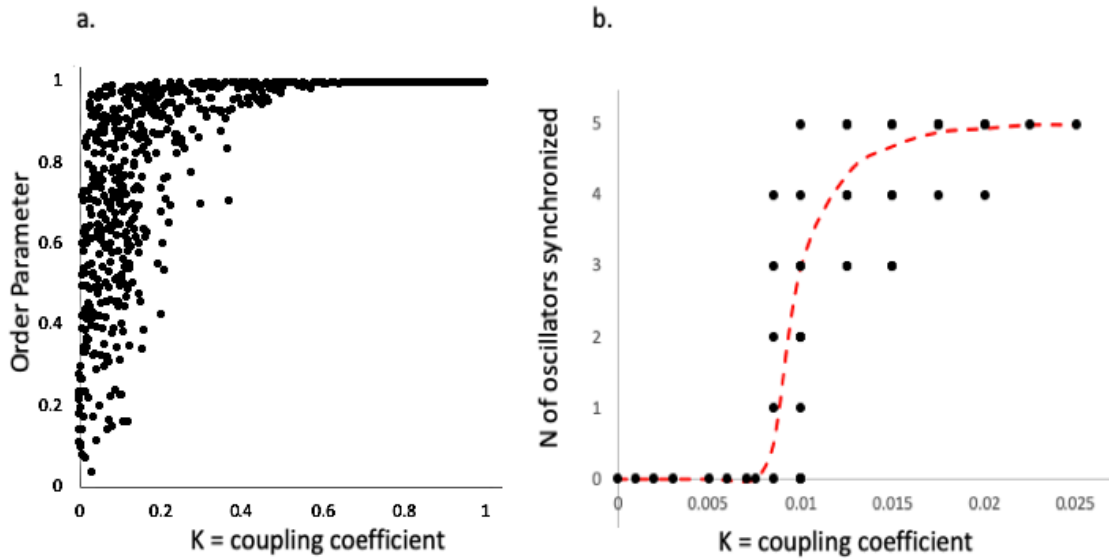

*Figure 2. Synchrony patterns as a function of coupling strength. a. The order parameter (equation 3) as a function of K. The values of K are scaled to the 22 oscillators in the system, which is to say the K in the graph is k/22 (from equation 2). Unusually high variance is due to the small number of oscillators and specific clustered coupling pattern. b. The coffee berry borer group only (oscillators 17-21) calculating how many of the 5 are in synchrony based on the critical angular difference ($c_{i,j}$) of 0.01. Note the typical failure to synchronize at low levels, not captured by the traditional order parameter due to the small number of oscillators. Dashed curve connects the means at each coupling coefficient.*

The amount of scatter in the order parameter before full synchronization reflects a complicated approach to that synchrony, based on the sequential synchronization of independent synchrony groups. In Figure 3 we illustrate the approach to complete synchronization for one exemplary run. The structure of the synchrony graph is based on $C = 0.01$, and we see a clear division of oscillators in three distinct sets in Figure 3a. It is notable that oscillator 3 does not synchronize with any of the three groups. To fully understand the significance of the graph, it is convenient to follow the trajectories of all the oscillators over time as in figure 3b. It is evident that all the oscillators eventually synchronize as they converge to a common oscillatory angle θ. Yet the pattern of synchronization is not a uniform coming together of all the oscillators. It is clear that synchrony groups form rapidly (about t=5) and subsequently the groups themselves synchronize, perhaps most clearly seen in the three diagrams in the complex plane (Figure. 3c) placed at approximately the same position as the time series of the angles representing the oscillators. Thus, the organization of synchrony groups is clearly a transient phenomenon, suggesting the time to synchrony as a potentially important characteristic of the synchrony group itself.

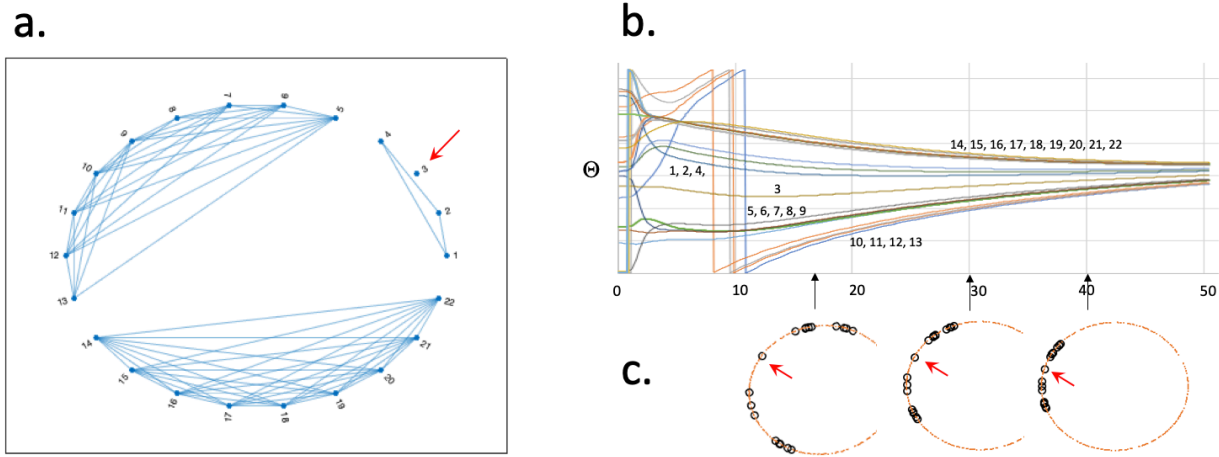

*Figure 3. Approach to synchrony groups and eventually to full synchronization for a single example. a. the graph of the synchrony groups obtained at time = 20 (connected nodes are in the same synchrony group). b. time series of Θ for all 22 oscillators, indicating which oscillator synchrony groups are formed. c. All 22 oscillators in the complex plane at three different points in the time series. In a and c, small red arrows point to oscillator number 3.*

In Figure 4 we illustrate the basic pattern of development of the stable synchrony groups as the coupling coefficient (K) is increased. Note that the coffee berry borer group, narrowly construed (i.e., oscillators 17 – 21) begins synchronizing at a very low value of K. Observations of the process reveal that any two of the five oscillators can synchronize first at K ~ 0.010, while any three of the five at K ~ 0.014, and any four of the five at K ~ 0.015. These small synchrony groups are not "stable" at these low coupling coefficients and likely reflect the random positioning of the oscillators at the beginning of a simulation. Similar observations are evident for the coffee leaf rust group (oscillators 10 – 13). It is clear, however, from Figure 4, that once a synchrony group is formed, it is invariant relative to further increases in K, with the further addition of other oscillators as K increases.

For K ~ 0.6, four distinct synchrony groups are stable. It is also notable that the makeup of the groups corresponds quite well to the general nature of the network with the four groups corresponding to the biological control system of each of the pests, another group associated with the phorid parasitoid, forms at a higher coupling coefficient. As coupling reaches approximately the level of 0.7, the coffee berry borer group merges with the leaf miner group, and the scale insect group merges with the coffee leaf rust fungus group at approximately K = 0.9. Oscillator 3, the predation of a spider on the small parasitoids, synchronizes only at the point that all oscillators are in synchrony.

It is also worth noting that over specific ranges of coupling, there are some oscillators that act something like oscillator 3, for a range of coupling values. For example, in the range K= 0.1 – 0.4, while there are two synchrony groups in which seven of the 22 oscillators are involved, while the remaining oscillators fail to synchronize either with one of these groups or among themselves. Such isolated oscillators are sometimes referred to as chimeric elements in the overall structural framework. Thus, when K = 0.6, for example, we could describe the "community structure" as consisting of four specific synchrony groups plus eight chimeric elements, at K = 0.7 we have three synchrony groups plus four chimeric elements, and at K = 0.8 we have four synchrony groups plus one chimeric element.

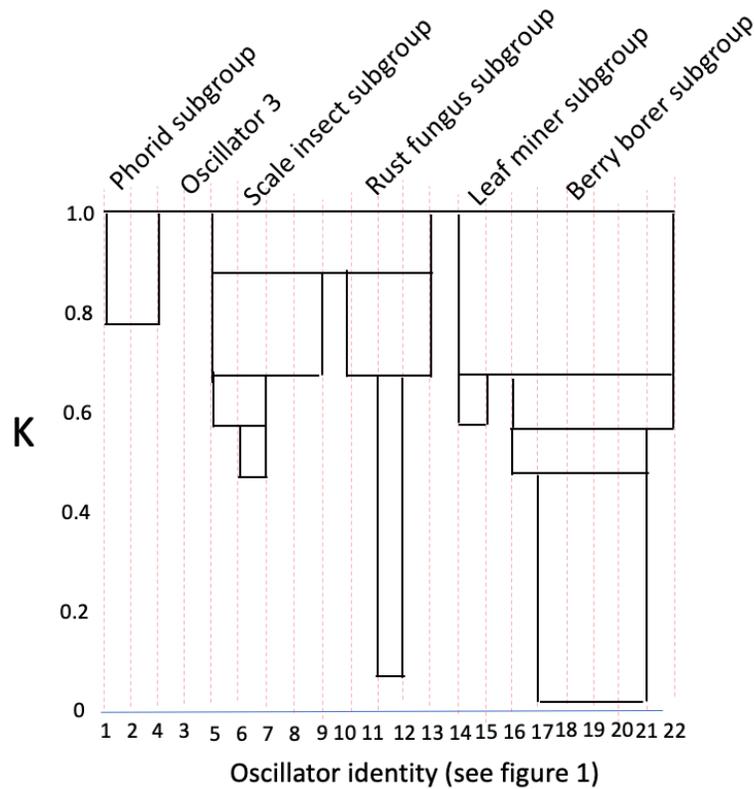

*Figure 4. Development of synchrony groups as a function of the coupling coefficient. The horizontal lines at various values of K show the onset of new synchrony groups, where oscillators share a rectangle when they form a synchrony group. The horizontal line at K = 1.0 indicates the entire system is synchronized.*

Thus, comparison of the model with this real-world ecological network reveals a complex set of patterns within the synchrony group range. Of particular interest is the fact that the larger synchrony groups are not necessarily predictable from a qualitative interpretation of the elementary network structure (Figure 1). For instance, oscillator 3 is coupled with both oscillator 8 and oscillator 2, and those two oscillators are in different large synchrony groups (i.e., when K is high), explaining why oscillator 3 only joins in synchrony when the large group synchronizes, since even at very high coupling strengths it apparently remains chimeric. Furthermore, the coupling of 8 with 3 connects what would otherwise be completely independent networks ({1-4, 14 – 22} and {5-13} – see inset in Figure 1). Thus, if oscillator 3 were eliminated from the system, the complete synchronization of the entire ensemble would not occur at all, even though that very oscillator is extremely resistant to synchronizing with other oscillators, because of its coupling to elements in the two major independent large synchrony groups.

An alternative look at the basic graph (Figure 1, inset) reveals another obvious division of the graph into independent subgraphs with the elimination of oscillator 22 (leaving the two completely separate groups 1-13 and 14-21). This is an interesting point of departure for examining the consequence of different weightings of the coupling coefficient, since this particular connection is indeed a bit special in the real system. Oscillator 22 involves the keystone ant species *Azteca seriaceasur* and its consumption of the coffee berry borer. There is some debate among experts (see for example the alternative interpretations of Vandermeer et al., 1997 versus Jiménez-Soto et al., 2013) as to how much energy transfer occurs in this interaction. Furthermore, there are dramatic trait-mediated indirect effects involved (Hsieh et al., 2012, Liere and Larson, Liere and Perfecto 2008) which ultimately must translate into at least weak coupling of 22 with all the other oscillators associated with the coffee berry borer. Consequently, we did a series of simulations setting the subset of coupling coefficients involving oscillator 22 to 10% less than all the other couplings. The results of this were that, on the one hand, the basic smaller synchrony groups emerged similarly as to when all coefficients were constant, suggesting that it is the structure of the network that mainly drives the overall community dynamics and structure, in the case with these adjustments to K. On the other hand, subtleties did emerge. For example, in our original analysis, oscillator 3 (an orb weaving spider, *Pocobletus* sp., a *Linephidae* catching parasitoids in its web) failed to synchronize with any group unless the coupling became very strong (Figure 4). Yet, after we lowered oscillator 22 couplings, it merged with one of the two major groups (1 - 13) at the relatively low general coupling of $K = 0.5$. Based on the real network structure (Figure 1), we infer that when oscillator 22 is tightly coupled with its associated pairs, oscillator 3 is pulled in both major synchrony group directions, first because of its "indirect" coupling with oscillator 22 (through the couplings with oscillators 2 and 4) pulling it in the direction of group 14 – 21 simultaneously with its coupling with oscillator 8, also pulling it in the direction of group 1-13. As the system becomes completely coupled, the major transition that previously occurred from four large synchrony groups to two (and eventually to complete synchrony) changes dramatically to include entirely different members when the coupling coefficient of oscillator 22 is small (Figure 5), again consistent with evidence of the importance of this ant-beetle interaction for the whole community and agro-ecosystem.

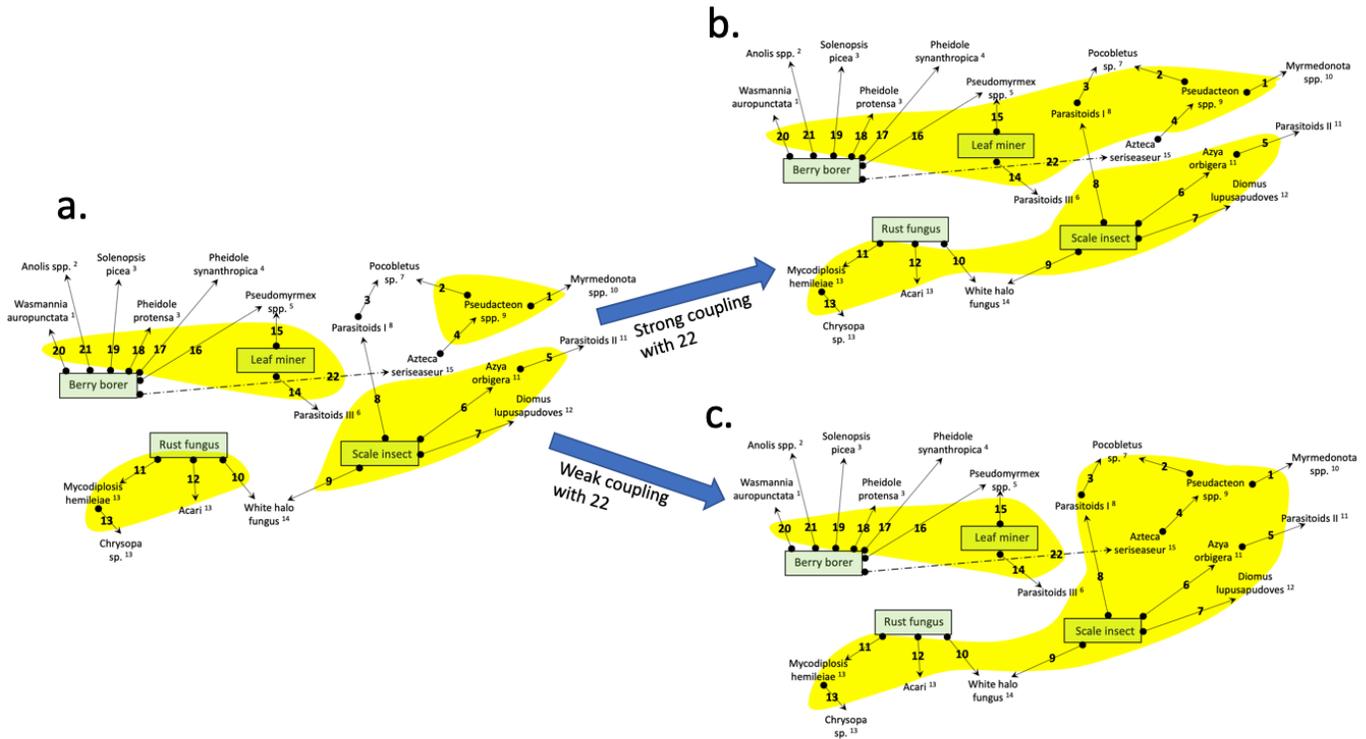

*Figure 5. Last two synchrony groups before complete coupling of the whole system, depending on whether the coupling of oscillator 22 to the system is weak or strong. a. the four synchrony groups before merging into a total of two groups. b. before final synchronization of all oscillators, with strong coupling with oscillator 22. c. before final synchronization of all oscillators, with weak coupling with oscillator 22. Note that, metaphorically, when oscillator 22 is weak, the joint coupling of oscillators 8 and 9 combine to create the mega group of the coffee leaf rust and scale groups with the phorid group, but when oscillator 22 is strong (equal coupling as the other oscillators), it dominates the tendency of either oscillators 8 or 9.*

**Discussion**

Reflecting on the prescient ideas of Platt and Denmann, we suggest that joining their perspective with the powerfully elegant model of Kuramoto (1975) provides a new window through which ecological communities and ecosystems might be examined both theoretically and empirically. Here we study this framework as applied to a real-world example of four pest species and their biological control elements, all of which are conceptualized as oscillators and coupled together in specific ways according to published studies. We argue that this mode of analysis generates new ways of looking at ecosystems. In particular, rather than a focus on species population densities (or biomasses) and attendant features such as diversity and traditional stability, it might be useful to analyze the system based on synchronization groups and patterns of synchronization. Such a suggestion depends on an initial assumption that the ecosystem is itself relatively permanent, which is to say the traditional questions of stability (Lyapunov or more generally) are not involved. Rather we look, theoretically, at patterns of approach to synchrony and the synchrony groups that emerge, and empirically, at spatial and temporal correlations predicted by the Kuramoto model (i.e., elements co-occurring in a synchrony group would likely be correlated empirically). Somewhat similar to the change in scale of description of groups of populations into meta-populations (Levins 1969), the attraction of our approach is that it permits prediction of coarser scale synchrony patterns to be expected based on nothing more than a qualitative understanding of consumption patterns. Indeed, in the particular example we analyzed herein, if the system presented in Figure 1 had been subjected to the 22 differential equations describing each species, and the minimal of two parameters associated with each of them, empirical parameter estimations would be unwieldy and, in our practical experience probably pointless since the values of those parameters clearly change over time and differ enormously from place to place. Yet, the existence of each of the oscillators is a clear reality, and the fact that they are very often dynamically coupled can hardly be questioned. In essence, the parameters of the system come directly from a knowledge of its natural history.

An issue that has attracted a great deal of attention in the literature surrounding the Kuramoto model is perhaps relevant to ecosystems as well, chimeric elements (Townsend et al., 2020; Strogatz, 2000; Abrams and Strogatz, 2004; Lee and Cho, 2019). The result herein that increasing coupling strength of the oscillators in the model may, under at least some circumstances, produce one or more synchrony groups, but also some individual oscillators that "refuse" to synchronize with any group at all. We refer to those as chimeric elements. In the most extreme case, a large number of identical oscillators synchronize at some critical coupling strength, but other oscillators, identical in every way, do not synchronize and continue wandering in state space, apparently in perpetuity. The idea of chimeric elements is similar in spirit to chaos in that it appears to emerge almost magically under some circumstances, and in physical systems also seems to have some empirical support (Tinsley et al., 2012; Kapitaniak et al., 2014). Mathematically there is some question about the reality of chimeric elements, although in practice extremely long transients are quite relevant in practical applications.

Given the extensive literature suggesting that chimeric elements, even if just long transients, exist in a wide variety of physical systems, we might also expect them to occur in ecological situations. If this be the case we can summarize further a vision of community structure as the structure of synchrony groups plus associated chimeric elements. So, for example, the synchrony groups that emerge in the present example are clearly associated with particular pest species, which is not surprising. The parallel question of how they relate to the natural history of the empirical system stems partly from "competition" from distinct synchrony groups. Thus, oscillator 3 in the present system, by bridging the gap between the two main synchrony groups (at relatively high values of K) is pulled in both directions and, at best, stabilizes in a position between them, but never moves toward one or another.

Yet the idea of a chimeric element perhaps recalls a basic idea in community ecology. It is easy to postulate other, perhaps more complicated, ecosystems, or perhaps with additional complications to the basic Kuramoto framing, as having chimeric elements in addition to synchrony groups. For example, the idea of fugitive species (Dayton, 1975; Horn and MacArthur, 1972), a common referent in early literature, fits this idea perfectly, in that we could imagine the normal semi-stable or permanent community, structured along the lines of synchrony groups, but the fugitive species coming and going, apparently without fitting into the basic community structure of the permanent resident species. In this sense the idea of a chimeric element parallels the well-known ecological idea of a fugitive species.

Our general results may be summarized as follows: First, as strength of oscillator couplings increases, synchrony groups tend to form, each one of which is associated with a particular pest species. Second, these groups tend to merge, ultimately forming three major groups, 1) the coffee berry borer/miner group, 2) the coffee leaf rust/scale insect group, and 3) the phorid group (Figures 3,4), all mirroring ecological structures within the system. Third, relaxing the assumption of perfectly uniform couplings, allowing the *A. seriaceasur*/CBB oscillator (number 22) to be only a tenth of the others (based on our knowledge of the system), provokes changes, the main one of which is that the phorid group synchronizes with the coffee leaf rust/scale group rather than the coffee berry borer/miner subgroup as it did previously. We conclude that the general pattern of synchrony group formation is a consequence of the structure of the network, but that particulars of synchrony group formations may vary depending on heterogeneity of oscillator coupling strengths. All of the synchrony group results that emerge from the Kuramoto model make perfectly consistent sense with what we know of the system in the field, an encouraging result.

It must be noted that this real world system itself is represented in a simplified fashion, as we know from the large amount of empirical work that has already been published (see references in caption to figure 1). Most important are the trait-mediated indirect interactions (or higher order interactions) associated, for example, with oscillator 22, which are known to be important empirically (Jimenez-Soto et. al. 2013), and, given the analysis here, are likely to be important in the further detailed study of synchrony groups of the system (Liere and Larson 2010; Liere and Perfecto, 2014). Incorporating such complexities is a challenge for the future.

Nevertheless, even with this simplified version of the real system, certain patterns suggest qualitative predictions. For example, the set of oscillators associated with the scale insect (5 - 8), is almost always a synchrony group, suggesting that spatial and temporal correlations among the members of that group should be observable in nature. Such a prediction is, in our practical experience in the field, certainly true, (Vandermeer and Perfecto, 2019; Iverson et al., 2018). Furthermore, our model of the system also provides new predictions about the spatiotemporal correlations that may be apparent given the modification of the underlying coupling of components of the system. This may be realized in the empirical agroecosystem through several avenues such as coupling strengths being modified due to seasonal factors or under different regimes of management of the agroecosystem.

The present study is in the spirit of many other network studies in framing the question "what can I say about the system knowing nothing more than which nodes are connected". Since the nodes we propose here are not species or populations but, rather, oscillators, and our focus is not stability or permanence but rather synchrony, perhaps some unique insights emerge. Connecting this theoretical program with empirical work provides, perhaps, more insight than a direct application of a dynamic model (e.g., a system of ODEs) to the system. The pattern of oscillator couplings (which oscillators are coupled with which), determination of which is a rather easy empirical exercise, suggests questions about qualitative patterns of co-occurrence in the field (which elements are expected to be correlated in their spatial and/or temporal occurrence) which would be subject to empirical verification.

**Acknowledgements**: This work was supported by NSF grant number DEB: 1853261.